 \documentclass[12pt]{article}

\usepackage{amsmath}

\begin{document}

\title{Generating potentials via difference equations}
\author{S. D. Maharaj\thanks{Author
for correspondence; email: maharaj@ukzn.ac.za} \; and S.
Thirukkanesh\thanks{Permanent address: Department of Mathematics,
Eastern University, Chenkalady, Sri Lanka} \\
Astrophysics and Cosmology Research Unit\\
School of Mathematical Sciences \\ University of KwaZulu-Natal\\
Durban 4041, South Africa}
\date{}
\maketitle

\begin{abstract}
The condition for pressure isotropy, for spherically symmetric
gravitational fields with charged and uncharged matter, is reduced
to a recurrence equation with variable, rational coefficients. This
difference equation is solved in general using mathematical
induction leading to an exact solution to the Einstein field
equations which extends the isotropic model of John and Maharaj. The
metric functions, energy density and pressure are well behaved which
suggests that this model could be used to describe a relativistic
sphere. The model admits a barotropic equation of state which
approximates a polytrope  close to the stellar centre.
\end{abstract}

AMS Classification numbers: 83C15; 83C22; 85A99

\section{Introduction}
Solutions of the Einstein field equations for spherically symmetric
gravitational fields in static  manifolds are necessary in the
description of compact objects in  relativistic astrophysics. The
models generated are utilised to describe relativistic compact
objects where the gravitational field is strong as is the case in
neutron stars. It is for this reason that considerable energy and
time is devoted to the study of the mathematical properties and
features of the underlying nonlinear differential equations. The
detailed lists of Stephani {\it et al} \cite{Stephani} and Delgaty
and Lake \cite{Lake} for static, spherically symmetric models
provide a comprehensive collection of interior spacetimes that match
to the Schwarzschild exterior spacetime. It is important to note
that only a few of these solutions correspond to nonsingular metric
functions with a physically acceptable energy momentum tensor. Some
of the exact solutions to the field equations, which satisfy all the
physical requirements for a relativistic  star,  are contained in
the models of the Durgapal and Bannerji \cite{Durgapal}, Durgapal
and Fuloria \cite{Fuloria}, Finch and Skea \cite{Finch}, Ivanov
\cite{Ivanov}, Maharaj and Leach \cite{Maharaj} and Sharma and
Mukherjee \cite{SharmaMukherjee2002}, amongst others.

In this paper our objective is to find new exact solutions to the
Einstein field equations which may be used to describe the interior
spacetime of a relativistic sphere. The approach essentially reduces
 to the analysis of difference equations which we demonstrate
leads to explicit solutions.
 We first express the Einstein equations, for neutral matter, as a
new set of differential equations utilising a transformation due to
Durgapal and Bannerji \cite{Durgapal} in \S2. We choose a general
polynomial form for one of the gravitational potentials, which we
believe has not been studied before. This enables us to simplify the
condition of pressure isotropy in \S3 to a second order linear
equation in the remaining gravitational potential. We assume a
series form for this function which yields a difference equation
which we manage to solve using mathematical induction. It is then
possible to exhibit a new exact solution to the Einstein field
equations which can be written explicitly as shown in \S4. Our
results contain the model of John and Maharaj \cite{John} as a
special case. The curvature and matter variables appear to be well
behaved in the interior spacetime. We also demonstrate the existence
of an explicit barotropic equation of state relating the pressure to
the energy density. For small values of the radial coordinate, close
to the stellar core, we demonstrate that the equation of state
approximates a polytrope. Then in \S5 we consider the
Einstein-Maxwell equations for charged matter. The solutions found
in the presence of an electromagnetic field  reduce to the model for
neutral matter given earlier in \S4. We believe that the method
illustrated in this paper is a useful device in the production of
models for compact objects.

\section{Field equations}
We  assume that the spacetime manifold is static and spherically
symmetric. This requirement is  consistent with models utilised to
study physical processes in relativistic astrophysical objects such
as dense stars. The generic line element for static, spherically
symmetric spacetimes is given by
    \begin{equation}
    ds^2 = -e^{2\nu(r)}dt^2 + e^{2\lambda(r)} dr^2
    + r^{2} (d\theta^2 +\sin^{2}\theta \,d\phi^{2}) \label{spherical}
    \end{equation}
in Schwarzschild coordinates. For neutral perfect fluids the
Einstein field equations can be expressed as follows
\begin{subequations} \label{perfect}
    \begin{eqnarray}
\frac{1}{r^2}  [r(1-e^{-2\lambda})]'& =& \rho \label{perfect-a}
\\   -\frac{1}{r^2}\ (1-e^{-2\lambda})+\frac{2\nu'}{r}\
e^{-2\lambda}& =& p \label{perfect-b}  \\   e^{-2\lambda}
\left(\nu''+{\nu}'^{2} +\frac{\nu'}{r}  -\nu' \lambda'-
\frac{\lambda'}{r}\right) & = & p \label{perfect-c}
\end{eqnarray}
    \end{subequations}
for the spherically symmetric line element (\ref{spherical}).
 The energy density $\rho$ and the pressure $p$ are measured relative
to the comoving fluid 4--velocity $u^a = e^{-\nu} \delta^{a}_{\,0}$
and primes denote differentiation with respect to the radial
coordinate $r.$ In the field equations (\ref{perfect}) we are
utilising units where the coupling constant $\frac{8 \pi G}{c^4} =
1$ and the speed of light  $c=1.$

A different but equivalent form of the field equations is obtained
if we introduce a new independent variable $x$, and new metric
functions $y$ and $Z$, as follows
\begin{equation}  A^2 y^2 (x) = e^{2 \nu(r)}, \,\,\, Z(x) =
e^{-2 \lambda(r)}, \,\,\, x = C r^2 \label{transf}
 \end{equation}
 In the transformation (\ref{transf}), due to
 Durgapal and Bannerji \cite{Durgapal}, the quantities $A$ and $C$ are
 arbitrary constants. Under
the transformation (\ref{transf}), the system (\ref{perfect}) has
the form
    \begin{subequations} \label{Finch}
    \begin{eqnarray}
    \frac{1-Z}{x}-2\dot{Z} &=&\frac{\rho}{C}            \label{Finch-a}\\
    4Z\frac{\dot{y}}{y} + \frac{Z-1}{x} &=& \frac{p}{C} \label{Finch-b}\\
    4Zx^2 \ddot{y} + 2\dot{Z}x^{2} \dot{y} + (\dot{Z}x-Z+1)y&=&0    \label{Finch-c}
    \end{eqnarray}
    \end{subequations}
where the dots denotes differentiation with respect to the variable
$x$. It is clear that (\ref{Finch}) is a system of three equations
in the four unknowns $\rho, p, y$ and $Z$. The advantage of using
(\ref{Finch}), rather than the original system (\ref{perfect}), is
that on a suitable specification of $Z(x)$, (\ref{Finch-c}) becomes
second order and linear in $y$.

\section{Difference equations}

The approach that we follow is to specify the gravitational
potential $Z(x)$ and attempt to solve (\ref{Finch-c}) for the
potential $y.$
 From inspection it is clear that the simplest solutions to the system
  (\ref{Finch}) correspond to polynomials forms for $Z(x)$.
  Consequently in an
attempt to obtain a new solution to the system (\ref{Finch}) we make
the particular choice
    \begin{equation}
Z = 1 + ax^{k}  \label{ae1}
    \end{equation}
where $a$ is a constant. As far
  as we are aware all exact solutions found previously
   correspond to forms of the gravitational potential $Z(x)$
   which are linear ($k=1$) or quadratic ($k=2$) in the independent variable $x$.
   For a comprehensive list of spherically symmetric solutions
   corresponding to different choices of $Z(x)$,
   relevant to relativistic astrophysics, see Delgaty and Lake
   \cite{Lake}
   and Stephani {\it et al} \cite{Stephani}.
Recently John and Maharaj \cite{John} generated a new solution
corresponding to the cubic form with $k=3$.  Higher values for $k$
have not been considered before because the resulting differential
equation in the dependent variable $y$ is difficult to solve.  With
the specified function $Z$  given above the condition of pressure
isotropy (\ref{Finch-c}) becomes
\begin{equation}
2 \left( 1+ax^{k} \right) \ddot{y} + akx^{k-1} \dot{y} + a \alpha
x^{k-2}y = 0 \label{cubic4}
\end{equation}
where we have set \(\alpha = \frac{k-1}{2} \) for convenience. The
linear second order differential equation (\ref{cubic4}) is
difficult to analyse when $a\not=0$. Standard handbooks of
differential equations and computer software packages such as
Mathematica have not been helpful in generating exact solutions to
(\ref{cubic4}). We attempt to find a series solution to
(\ref{cubic4}) using the method of Frobenius. As the point $x=0$ is
a regular point (for $k>2$) of (\ref{cubic4}), there exist two
linearly independent solutions of the form of a power series with
centre $x=0$. Therefore it is possible to write
    \begin{equation}
y(x) = \sum^{\infty}_{n=0}c_{n} x^{n}   \label{cubic5}
    \end{equation}
where the $c_n$ are the coefficients of the series. For an
acceptable solution we need to determine the coefficients $c_n$
explicitly.

Substituting (\ref{cubic5})  into (\ref{cubic4}) yields
    \begin{eqnarray*}
2\sum_{n=2}^{k-1}c_{n}n(n-1)x^{n-2} + [2c_{k}k(k-1)+
a\alpha c_{0}]x^{k-2}     &   &  \\
+ [2c_{k+1}(k+1)k + ac_{1} (k + \alpha)]x^{k-1} &   &  \\
 + \sum_{n=2}^{\infty} \left[ 2c_{n+k}(n+k)(n+k-1)
+ ac_{n}(2n(n-1) + kn + \alpha)  \right]x^{n+k-2} & = & 0.
\end{eqnarray*}
Note that in this equation each term contains ascending powers of
$x$. For this result to hold true for all $x$ we require that the
coefficients satisfy a set of consistency conditions
 \begin{subequations} \label{cubic9}
\begin{eqnarray}
c_{n} & = & 0 , ~ n = 2,3, .. ,k-1 \label{cubic9-a} \\
2c_{k}k(k-1) + a\alpha c_{0}      & = & 0 \label{cubic9-b} \\
2c_{k+1}(k+1)k +ac_{1} (k+\alpha) & = & 0   \label{cubic9-c}\\
  2c_{n+k}(n+k)(n+k-1)+ ac_{n}[2n(n-1) + kn + \alpha]
 & = & 0 , ~ n  \geq  2. \label{cubic9-d}
\end{eqnarray}
\end{subequations}
The recurrence relation (\ref{cubic9-d}) is of order $k$ and
consists of variable, rational coefficents. Equation
(\ref{cubic9-d}) does not fall in the known classes of difference
equations and needs to be analysed from first principles.
 Note that (\ref{cubic9-a}) and
(\ref{cubic9-d}) imply
\begin{equation}
c_{k+2} = c_{k+3} = \cdots = c_{k+(k-1)}=0 \label{zero}
\end{equation}
Hence the remaining nonzero terms are applicable from $n \geq k$ in
(\ref{cubic9-d}). Consequently the system (\ref{cubic9}) is reduced
to the following set
\begin{subequations} \label{cubic10}
\begin{eqnarray}
c_{k}   & = & -\frac{a}{2} \frac{\alpha}{k(k-1)}c_{0} \label{cubic10-a} \\
c_{k+1} & = & -\frac{a}{2} \frac{(k+\alpha)}{(k+1)k} c_{1} \label{cubic10-b} \\
c_{n+k} & = & -\frac{a}{2} \left[ \frac{2n(n-1) + kn +
\alpha}{(n+k)(n+k-1)} \right] c_{n} ,~n\geq k. \label{cubic10-c}
\end{eqnarray}
\end{subequations}
Note that from (\ref{cubic9-a}) and (\ref{zero}) we have two sets of
consecutive
 coefficients which vanish. This pattern of zero coefficients
repeats itself because of the recurrence relation (\ref{cubic10-c}).
 The nonvanishing coefficients can be placed into two groups.
These can be written either in terms of the first leading
coefficient $c_0$ or in terms of the second leading coefficient
$c_1$. We consider these sets in turn.

 We first consider the coefficients
\[c_0, c_k, c_{2k}, c_{3k}, \dots  \]
From the system (\ref{cubic10}) we can generate expressions for
$c_k, c_{2k}, c_{3k}, \dots$ in terms of the first leading
coefficient $c_0$. These coefficients generate a pattern and we can
write
\begin{equation}
c_{nk+k}=\left(- \frac{a}{2}\right)^{n+1}\prod_{p=0}^{n}
 \frac{[2(kp)(kp-1)+k(kp)
+\alpha]}{(kp+k)(kp+k-1)} c_{0}  , \,\,\, n\geq 0
\label{difference1}
\end{equation}
where we have utilised the conventional symbol $\prod$ for
multiplication. It is also possible to establish the result
(\ref{difference1})
 rigorously applying mathematical induction. For $n=0$
 the result (\ref{difference1}) is obvious since
\[
c_{k}
         =   \left(- \frac{a}{2}\right)^{0+1}\frac{2 .0 (0-1)
+k . 0 + \alpha}{(0+k)(0+k-1)}c_{0}
\]
Now suppose that the result (\ref{difference1}) holds for $n=r$
which is the inductive step so that we can write
\[ c_{rk+k}=\left(- \frac{a}{2}\right)^{r+1}
\prod_{p=0}^{r} \frac{[2(kp)(kp-1) +k(kp) +\alpha]}{(kp+k)(kp+k-1)}
c_{0}. \] Then let $n=r+1$ in (\ref{difference1}) which is the next
term.
 From equation (\ref{cubic10-c}) we have that
\[
c_{(r+1)k+k}  =
  - \frac{a}{2}\frac{[2k(r+1)(k(r+1)-1)
 + k(k(r+1))+\alpha]}{(k(r+1)+k)(k(r+1)+k-1)}c_{rk+k}
\]
By the above inductive step this is
 equivalent to
\begin{eqnarray*}
c_{(r+1)k+k} & = &  - \frac{a}{2}\frac{[2k(r+1)(k(r+1)-1)
 + k(k(r+1))+\alpha]}{(k(r+1)+k)(k(r+1)+k-1)} \times  \\
&   & \left(- \frac{a}{2}\right)^{r+1}\prod_{p=0}^{r}
 \frac{[2(kp)(kp-1)+k(kp)
+\alpha]}{(kp+k)(kp+k-1)} c_{0}  \\
 & = & \left(- \frac{a}{2}\right)^{r+2}\prod_{p=0}^{r+1}
 \frac{[2(kp)(kp-1)+k(kp)
+\alpha]}{(kp+k)(kp+k-1)} c_{0}.
\end{eqnarray*}
The above equation shows that the result  holds
 for $n=r+1$. Hence by the principle of mathematical
induction the result (\ref{difference1}) is true for all nonnegative
integers $n$.

 We can perform a similar analysis for
the coefficients
\[c_1, c_{k+1}, c_{2k+1}, c_{3k+1}, \dots \]
and attempt to establish a general pattern. From the system
(\ref{cubic10}) we can obtain the expressions for $c_{k+1},
c_{2k+1}, c_{3k+1}, \dots$ in terms of the second leading
coefficient $c_1$. These coefficients also generate a similar
pattern as above and we can write
\begin{equation}
c_{nk+k+1}  = \left(- \frac{a}{2}\right)^{n+1}
\prod_{p=0}^{n}\frac{[2(kp+1)(kp)
+k(kp+1)+\alpha]}{(kp+k+1)(kp+k)}c_{1}, \,\,\, n \geq 0
\label{difference2}
\end{equation}
and $\prod$ denotes multiplication. As in the previous case it is
possible to verify the result (\ref{difference2}) rigorously.  For
$n=0$ the result (\ref{difference2}) is obvious since
\[
c_{k+1}
           =  \left(- \frac{a}{2}\right)^{0+1}\frac{[2(k.0+1)(k.0)+k(k.0+1)
+\alpha]}{(k.0+k+1)(k.0+k)}c_{1} \]
 Now suppose
that the result (\ref{difference2}) holds for $n=r$. Then we have
\[ c_{rk+k+1} = \left(- \frac{a}{2}\right)^{r+1}
\prod_{p=0}^{r}\frac{[2(kp+1)(kp)+k(kp+1)
+\alpha]}{(kp+k+1)(kp+k)}c_{1}. \] Now let $n=r+1$ in
(\ref{difference2}) which is the next term.
 From equation (\ref{cubic10-c}) we have
\[ c_{(r+1)k+k+1} = - \frac{a}{2} \frac{2[(r+1)k+1](r+1)k
+ k[(r+1)k+1]+\alpha}{[k(r+1)+k+1][k(r+1)+k]} c_{(r+1)k+1}. \] By
the above inductive step this is equivalent to
\begin{eqnarray*}
c_{(r+1)k+k+1} & = & - \frac{a}{2} \frac{2[(r+1)k+1](r+1)k
+ k[(r+1)k+1]+\alpha}{[k(r+1)+k+1][k(r+1)+k]} \times  \\
 &   & \left(- \frac{a}{2}\right)^{r+1}\prod_{p=0}^{r}
\frac{[2(kp+1)(kp)+k(kp+1) +\alpha]}{(kp+k+1)(kp+k)}c_{1}  \\
 & = & \left(- \frac{a}{2}\right)^{(r+1)+1}
\prod_{p=0}^{r+1}\frac{[2(kp+1)(kp)+k(kp+1)
+\alpha]}{(kp+k+1)(kp+k)}c_{1}.
\end{eqnarray*}
The above equation shows that the result (\ref{difference2})
 holds for $n=r+1$. Hence by the principle of
mathematical induction the result (\ref{difference2}) holds for all
nonnegative integers $n$.

 Thus we have solved the general
recurrence relation in (\ref{cubic9}) and (\ref{cubic10}). The
coefficients  $ c_{k}, c_{2k}, c_{3k}, \dots $  are generated from
(\ref{difference1}). The coefficients  $ c_{k+1}, c_{2k+1},
c_{3k+1}, \dots $  are generated from (\ref{difference2}). Hence the
difference equation
 (\ref{cubic10-c})) has been solved and all nonzero coefficients are
 expressible in terms of the leading coefficients $c_{0}$
 and $c_{1}$.
 From the equations (\ref{cubic5}), (\ref{difference1}) and
 (\ref{difference2}) we can write the general
solution to (\ref{cubic4}) as
\begin{eqnarray}
y & = & c_{0} + c_{1}x + c_{k}x^{k} +c_{k+1}x^{k+1}+c_{2k}x^{2k}
+c_{2k+1}x^{2k+1}+ \cdots \nonumber \\
  & = & c_{0}\left[ 1 + \sum_{n=0}^{\infty}
 \left(- \frac{a}{2}\right)^{n+1}\prod_{p=0}^{n}
 \left(\frac{[2(kp)(kp-1)+k(kp)
+\alpha]}{(kp+k)(kp+k-1)} \right)x^{kn+k} \right]+ \nonumber \\
  &   & c_{1} \left[ x + \sum_{n=0}^{\infty}
\left(- \frac{a}{2}\right)^{n+1}\prod_{p=0}^{n}
 \left( \frac{[2(kp+1)(kp)+k(kp+1)
+\alpha]}{(kp+k+1)(kp+k)}\right) x^{kn+k+1} \right] \label{cubic13}
\end {eqnarray}
where $c_{0}$ and $c_{1}$ are arbitrary constants and $\alpha =
\frac{k-1}{2}$. Clearly the solution (\ref{cubic13}) is of the form
\begin{equation}
y(x) = c_{0}y_{1}(x) + c_{1}y_{2}(x) \label{cubic14}
\end{equation}
where
\begin{subequations} \label{cubic15}
\begin{eqnarray}
y_{1} & = & 1 + \sum_{n=0}^{\infty} \left(-
\frac{a}{2}\right)^{n+1}\prod_{p=0}^{n}\left(
\frac{[2(kp)(kp-1)+k(kp)
 +\alpha]}{(kp+k)(kp+k-1)} \right)x^{kn+k}  \\
y_{2} & = & x+
\sum_{n=0}^{\infty}\left(-\frac{a}{2}\right)^{n+1}\prod_{p=0}^{n}
\left(
\frac{[2(kp+1)(kp)+k(kp+1)+\alpha]}{(kp+k+1)(kp+k)}\right)x^{kn+k+1}
\end{eqnarray}
\end{subequations}
are linearly independent functions. Therefore we have found the
general solution to the differential equation (\ref{cubic4}) for the
particular gravitational potential $Z$ given in (\ref{ae1})

From (\ref{cubic14}) and (\ref{cubic15}) we can generate a number of
new particular exact solutions for specified values of $k$ and $a$.
For certain values of $k$  the solution  may reduce to models that
have already been found. We let $k=3$ so that $\alpha = 1$. Then the
gravitational potential $Z$ becomes
\[
Z = 1 + ax^{3}
\]
and the gravitational potential $y$ is
\begin{eqnarray}
y & = & c_{0}\left[ 1 + \sum_{n=0}^{\infty} \left(-
\frac{a}{2}\right)^{n+1}\prod_{p=0}^{n}\left( \frac{[2(3p)^{2}
 +3p +1 ]}{(3p+3)(3p+2)} \right)x^{3n+3} \right]+ \nonumber \\
  &   & c_{1} \left[ x + \sum_{n=0}^{\infty}
\left(- \frac{a}{2}\right)^{n+1}\prod_{p=0}^{n}
 \left( \frac{[2(3p+1)^{2}+(3p+1)+1]}{(3p+4)(3p+3)}\right) x^{3n+4} \right].
 \label{John}
\end {eqnarray}
Therefore our general solution contains as a special case
(\ref{John}) which is the model of John and Maharaj \cite{John}.

\section{Einstein models}

From the analytic representation (\ref{cubic14}), (\ref{cubic15})
and the
 Einstein field equations (\ref{Finch}) we generate the exact solution
\begin{subequations} \label{cubic16}
\begin{eqnarray}
e^{2\lambda}   & = & \frac{1}{1+ax^{k}} \label{cubic16-a}\\
e^{2\nu}       & = & A^{2}y^{2} \label{cubic16-b} \\
\frac{\rho}{C} & = & -a(1+2k)x^{k-1}  \label{cubic16-c}\\
\frac{p}{C}    & = & 4(1+ax^{k})\frac{\dot{y}}{y}+ax^{k-1}
\label{cubic16-d}
\end{eqnarray}
\end{subequations}
where $A$ and $C$ are constants, $y$ is given by (\ref{cubic14}),
 $a$ is constant and $k$ is a natural number greater than
one.  This solution has a simple form and is expressed completely in
terms of elementary functions and the series ({\ref{John}). The
metric functions and the matter variables are written explicitly in
terms of the independent variable $x$.

 The gravitational potentials
$\nu$ and $\lambda$ satisfy the conditions of physical reasonability
for a relativistic sphere: they are finite at the centre $x=0$,
continuous in the interior of the spacetime and match smoothly to
the Schwarzschild exterior spacetime at the boundary of the sphere.
To obtain positive energy density we require $a < 0 .$ The matter
variables $\rho$ and $p$ are bounded and nonsingular at the origin.
Even though our solution is presented in terms of an infinite
series, we point out that by using
 computer software packages it is easy to generate graphical
 plots of $\nu, \lambda, \rho$ and $p$. Thus a physical
analysis of our model is very feasible and this is an avenue for
future research. The expressions given above for the gravitational
potentials
 and the matter variables have the advantage of simplifying
 the analysis of the physical
features of the solution, and will assist in the description of a
relativistic compact bodies such as neutron stars.

From (\ref{cubic16}) we can observe that
\[x = \left({\frac{\rho}{-aC(1+2k)}}\right)^{\frac{1}{k-1}}, \,\,\, a< 0.    \]
Therefore from equation (\ref{cubic16-d}), we can express $p$ in
terms of $\rho$ only. This is a special feature of our model and
most of the solutions presented in the literature do not satisfy
this property. As this property does not depend on a particular
value of $k$, it is valid for allowable values of  $k$. Hence the
solution satisfies the barotropic equation of state
 \[p=p(\rho). \]
This property is usually made as a requirement for a physically
relevant model.  Also observe that for small values of $x$ close to
the stellar centre we have
 $y \approx c_0 + c_1x$. Then from (\ref{cubic16})
 we have the approximation
 \begin{equation}
 \frac{p}{C} \approx \frac{4 c_1}{c_0 + c_1
 \left[\frac{\rho}{-aC(1+2k)}
 \right]^{1/(k-1)}}
 \label{polytropic}
 \end{equation}
Therefore  for small values of $x$ close to stellar centre
(\ref{polytropic}) implies that we have the approximate equation of
state
\[ p \propto \rho^{-1/(k-1)} \]
which is of  the form of a polytrope. We make the observation that
equations of state, with pressure as a negative power of energy
density, arise in relativistic cosmological models with a Chaplygin
gas; these models are now widely used as a driver of acceleration to
generate the present universe \cite{Gas1,Gas2}.

\section{Einstein-Maxwell models}

It is possible to extend the solutions presented in this
 paper to include the electromagnetic field as the
procedure to obtain the new general solution is very similar to \S
3. We present only an outline and do not give all the details of the
calculation. A generalisation of the Einstein field equations is
given by
\begin{subequations} \label{cubic17}
\begin{eqnarray}
\frac{1-Z}{x} - 2\dot{Z} & = & \frac{\rho}{C} +
 \frac{E^{2}}{2C} \label{cubic17-a}\\
 4Z\frac{\dot{y}}{y} + \frac{Z-1}{x} & = & \frac{p}{C} -
 \frac{E^{2}}{2C}\label{cubic17-b} \\
 4Zx^{2}\ddot{y} + 2 \dot{Z}x^{2} \dot{y} + \left(\dot{Z}x -
Z + 1 - \frac{E^{2}x}{C}\right)y & = & 0 \label{cubic17-c}\\
 \frac{\sigma^{2}}{C} & = & \frac{4Z}{x} \left(x \dot{E} + E
\right)^{2} \label{cubic17-d}
\end{eqnarray}
\end{subequations}
where $E$ is the electric field intensity and $\sigma$ is the charge
density. When the electric field $E=0$ then the Einstein-Maxwell
equations (\ref{cubic17}) reduce to the Einstein equations
({\ref{Finch}) for neutral matter. The system of equations
(\ref{cubic17}) governs the behaviour
 of the gravitational field for a charged perfect fluid.

 On substituting (\ref{ae1}) in equation (\ref{cubic17-c}) we obtain
\begin{equation}
2 \left( 1+ax^{k} \right) \ddot{y} + akx^{k-1} \dot{y}
 + \left( a \frac{k-1}{2}x^{k-2} - \frac{E^2}{2Cx} \right)y = 0.
 \label{cubiciso}
\end{equation}
If we specify the electric field intensity to be
\[E^2 = 2a C \beta x^{k-1}\]
where $\beta$ is a constant, then (\ref{cubiciso}) becomes
\begin{equation}
2 \left( 1+ax^{k} \right) \ddot{y} + akx^{k-1} \dot{y}
 +  a (\alpha - \beta)x^{k-2}y = 0 \label{cubic18}
\end{equation}
where $\alpha = \frac{k-1}{2}$. This choice for $E$ is physically
reasonable as it is finite at the centre and continuous in the
interior spacetime. It is now possible to use the results of \S3 to
solve (\ref{cubic18}) directly. On comparing the differential
equations (\ref{cubic4}) and (\ref{cubic18}), we can write the
solution to the differential equation (\ref{cubic18}) as
\begin{equation}
y(x) = c_{0}y_{1}(x) + c_{1}y_{2}(x) \label{cubic19}
\end{equation}
where
\begin{subequations} \label{cubic20}
\begin{eqnarray}
y_{1} & = & 1 + \sum_{n=0}^{\infty} \left(-
\frac{a}{2}\right)^{n+1}\prod_{p=0}^{n}\left(
\frac{[2(kp)(kp-1)+k(kp)
+(\alpha -\beta)]}{(kp+k)(kp+k-1)} \right)x^{kn+k}  \\
      &   & \nonumber \\
y_{2} & = & x+ \sum_{n=0}^{\infty}\left(-\frac{a}{2}\right)^{n+1}
\prod_{p=0}^{n} \left( \frac{[2(kp+1)(kp)+k(kp+1)
+(\alpha - \beta)]}{(kp+k+1)(kp+k)}\right)x^{kn+k+1}. \nonumber \\
      &   &
\end{eqnarray}
\end{subequations}
From the analytic representation (\ref{cubic19}), (\ref{cubic20})
and the
 Einstein-Maxwell field equations (\ref{cubic17}) we generate the
new exact solution
\begin{subequations} \label{cubic21}
\begin{eqnarray}
e^{2\lambda}   & = & \frac{1}{1+ax^{k}} \\
e^{2\nu}       & = & A^{2}y^{2} \\
\frac{\rho}{C} & = & -a(1+2k+\beta)x^{k-1} \\
\frac{p}{C}    & = & 4(1+ax^{k})\frac{\dot{y}}{y}+a(1+\beta)x^{k-1} \\
\frac{E^2}{C}  & = & 2a\beta x^{k-1}
\end{eqnarray}
\end{subequations}
where $A, C, a$, and $\beta$ are constants, $y$ is given by
(\ref{cubic19}) and $k$ is a natural number greater than one. For
positive energy density we require $a<0$. We note that we can
express $p$
 in terms of $\rho$ only; hence this model also
satisfies the barotropic equation of state $p=p(\rho)$ as in the
case of neutral fluids. When $\beta = 0$, (\ref{cubic21}) reduces to
(\ref{cubic16}) which is the uncharged solution found earlier.

\section*{Acknowledgements}
ST thanks South Eastern University for study leave and the
University of KwaZulu-Natal for a scholarship. SDM and ST thank the
National Research Foundation of South Africa for financial support.

\end{document}